\documentclass[12pt]{iopart}
\usepackage{graphicx}
\usepackage{xcolor}

\usepackage{comment}
\usepackage{siunitx}
\usepackage[normalem]{ulem}
\expandafter\let\csname equation*\endcsname\relax
\expandafter\let\csname endequation*\endcsname\relax
\usepackage{amsmath}
\usepackage{amssymb}
\usepackage{braket}
\usepackage[colorlinks=False]{hyperref}
\begin{document}
\title[Feshbach Spectroscopy of Cs Atom Pairs in Optical Tweezers]{Feshbach Spectroscopy of Cs Atom Pairs in Optical Tweezers}
\author{R V Brooks$^1$\footnote{These two authors contributed equally.\label{note1}}, A Guttridge$^1$\footnotemark[0], Matthew D Frye$^2$, D K Ruttley$^1$, S Spence$^1$, Jeremy M Hutson$^2$ and Simon L Cornish$^1$}

\address{$^1$ Department of Physics and Joint Quantum Centre (JQC) Durham-Newcastle, Durham University, South Road, Durham DH1 3LE, United Kingdom

$^2$ Department of Chemistry and Joint Quantum Centre (JQC) Durham-Newcastle, Durham University, South Road, Durham DH1 3LE, United Kingdom
}
\ead{s.l.cornish@durham.ac.uk}

\begin{abstract}
\noindent
We prepare pairs of $^{133}$Cs atoms in a single optical tweezer and perform Feshbach spectroscopy for collisions of atoms in the states $(f=3, m_f=\pm3)$. 
We detect enhancements in pair loss  using a detection scheme where the optical tweezers are repeatedly subdivided. For atoms in the state $(3,-3)$, we identify  resonant features by performing inelastic loss spectroscopy. We carry out coupled-channel scattering calculations and show that at typical experimental temperatures the loss features are mostly centred on zeroes in the scattering length, rather than resonance centres.
We measure the number of atoms remaining after a collision, elucidating how the different loss processes are influenced by the tweezer depth. These measurements probe the energy released during an inelastic collision, and thus give information on the states of the collision products. We also identify resonances with atom pairs prepared in the absolute ground state $(f=3, m_f=3)$, where two-body radiative loss is engineered by an excitation laser blue-detuned from the Cs D$_2$ line. These results demonstrate optical tweezers to be a versatile tool to study two-body collisions with number-resolved detection sensitivity.

\end{abstract}


\section{Introduction}

Optical tweezers are a powerful tool for the study of ultracold collisions. They allow the preparation of an exact number of collision participants, with exquisite control of their internal states and sensitive number-resolved detection \cite{Tuchendler2008, Serwane2011, Xu2015, Liu2018, Sompet2019, Cheuk2020, Reynolds2020, Brooks2021}. This control may be combined with the ability to manipulate the collisional properties by tuning a magnetic field in the vicinity of a Feshbach resonance \cite{Chin2010}. Feshbach resonances underpin much of atomic physics; for example they are instrumental in the study of Bose-Einstein condensation (BEC) \cite{Leggett2001} and the production of ultracold molecules by magnetoassociation \cite{Kohler2006}. As such, use of Feshbach resonances is ubiquitous, and their detection and characterisation is of great interest \cite{Kohler2006,Chin2010}. 

Tweezers provide an excellent environment for studying the two-body physics of Feshbach resonances since three-body effects \cite{Weber2003b, Kraemer2006} are entirely suppressed by preparing exactly two atoms in a tweezer. The high densities achievable when there are two particles in the same tightly confining trapping potential aid the detection of extremely narrow resonances \cite{Mark2018}. It has also been proposed that a double-well tweezer could be used to estimate the pole strength of Feshbach resonances \cite{Jachymski2020}. By extension, effects involving three or more bodies could be measured by preparing exactly three or more atoms in the same tweezer \cite{Reynolds2020}.

Most experimental studies of ultracold collisions and Feshbach resonances have been performed in large-volume optical dipole traps containing many atoms \cite{Chin2010}. To reach the high particle densities required to observe ultracold collisions, evaporative cooling is often employed \cite{Ketterle1996}. However, efficient evaporative cooling imposes certain requirements on the starting densities and the collision properties of the trapped sample. The tight confinement of optical tweezers offers an alternative approach for the preparation of pairs of particles at sufficiently high densities to study collisions. This approach is particularly applicable to molecular species that can be laser cooled, such as CaF, which has recently been trapped in tweezers \cite{Anderegg2019,Cheuk2020}. So far, the detection of Feshbach resonances in optical tweezers has been limited to intraspecies resonances in fermionic $^6$Li \cite{Sala2013} and interspecies resonances in $^{23}$Na+$^{133}$Cs \cite{Hood2020,Zhang2020}. 

In this work, we study the intraspecies Feshbach resonances of pairs of bosonic \mbox{caesium-133} (Cs) atoms prepared in a single optical tweezer. Feshbach resonances in the $m_f=\pm3$ states of Cs have been extensively studied in bulk gases \cite{Vuletic1999b,Vuletic1999,Chin2000,Kerman2001,Chin2003,Herbig2003,Chin2004,Koppinger2014b}.
This allows us to benchmark our measurements in optical tweezers. We use inelastic loss spectroscopy to study Feshbach resonances in the $(f=3, m_f=-3)$ state. The inherent single-particle detection sensitivity of optical tweezers allows us to measure the number of particles remaining after a collision event. 
We utilise this to probe the rates of $2 \to 1$ and $2 \to 0$ atom loss as a function of tweezer depth. From these measurements we infer the energy released in inelastic collisions, finding good agreement with our expectations. We compare our loss measurements to coupled-channel scattering calculations. We find that, even for moderately low collision energies ($E_\textrm{coll}/k_\mathrm{B} \approx 2~\mu$K) that are typical for tweezer-based experiments, most of our observed loss features appear near the zeroes in the real part of the scattering length, not at the centres of the resonances, which correspond to peaks in the imaginary part of the scattering length. 
Finally, we perform Feshbach spectroscopy for atom pairs prepared in the $(f=3, m_f=+3)$ state. Since there are no inelastic channels in this case, we use radiative loss spectroscopy \cite{Vuletic1999} to observe multiple Feshbach resonances between 14 and 54~G.
 
The structure of the paper is as follows. In sections~\ref{subsec:prep} and \ref{subsec:transfer}, we describe the preparation of pairs of Cs atoms in a single optical tweezer. In sections~\ref{subsec:imaging} and \ref{subsec:detection}, we describe imaging of homonuclear atom pairs and detection of Feshbach resonances. In section~\ref{subsec:3-3}, we present the results of inelastic loss spectroscopy using the $(3,-3)$ state.
In section~\ref{subsec:radiative} we investigate Feshbach resonances in the (3,+3) state using radiative loss spectroscopy. In section~\ref{sec:theory}, we describe our coupled-channel calculations of the loss spectra, and discuss the origin of the loss features which appear at zeroes in the real part of the scattering length. In section~\ref{sec:depth}, we explore how the observed loss from the $(3,-3)$ state varies with trap depth. Section~\ref{sec:conclusion} concludes the paper.

\section{Experimental Methods}
\label{sec:experimental_methods}

\begin{figure}
    \centering
    \includegraphics[width=0.9\linewidth]{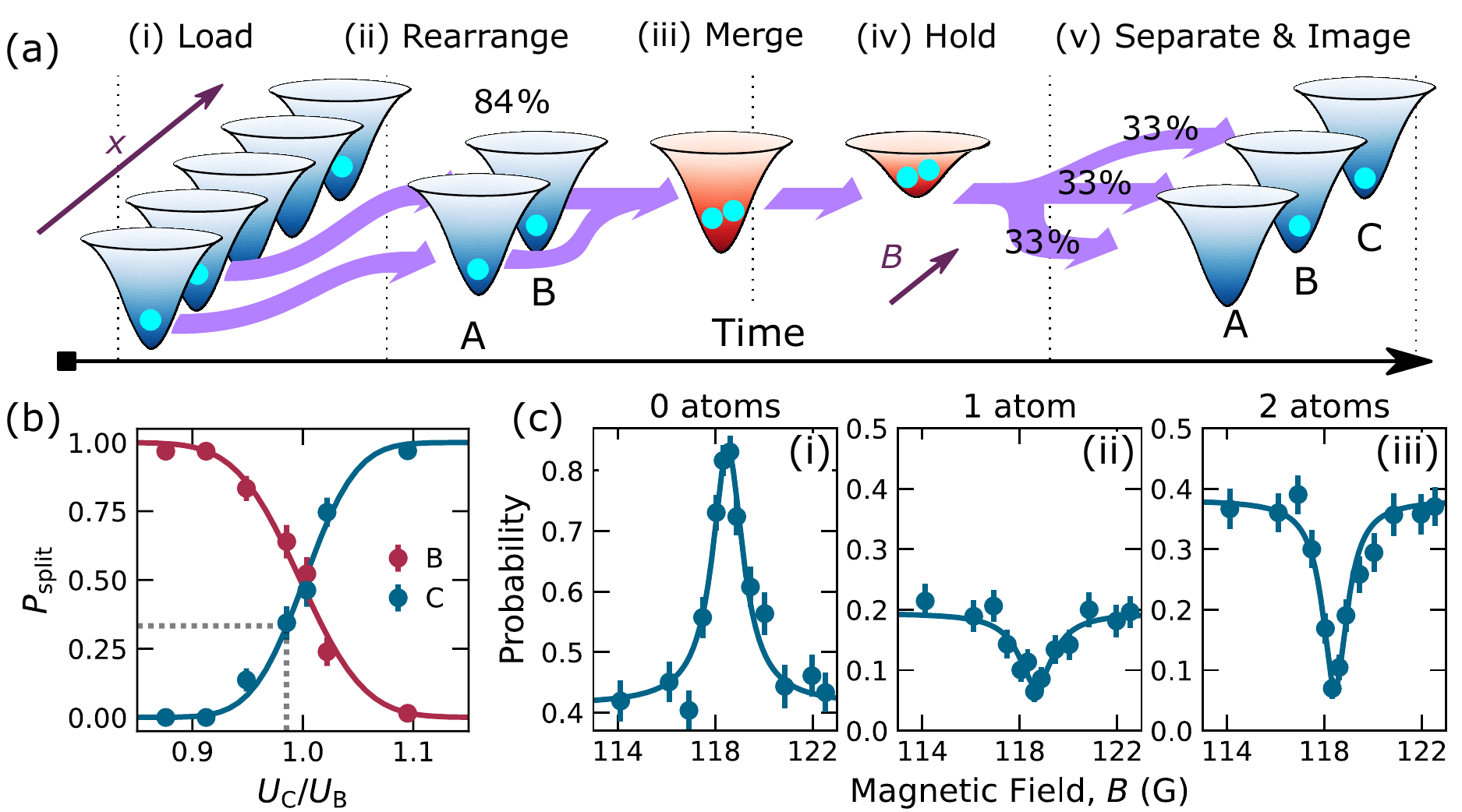}
    \caption{Preparation of atom pairs and detection of inelastic loss features. (a) Five tweezers of wavelength 938~nm are loaded, with a mean probability of 0.53(1) per tweezer. Following tweezer rearrangement along axis $x$ to sites A and B, pairs of Cs atoms are prepared in $84(2)~\%$ of shots. Cs pairs are merged into a 1064~nm tweezer ``collision tweezer'' (red), which is ramped to a depth of 30~$\mu$K for a hold time $t_\mathrm{coll}$ at a magnetic field $B$, during which there is a probability of pair loss. The survival is probed by separating the atoms into 3 final tweezers where they are imaged. Tweezer C is split first from tweezer B with a transfer probability of 33~$\%$. Tweezer A is then split from tweezer B with a 50~$\%$ transfer probability. This yields an approximately equal probability for a given atom to occupy any of the imaging traps. (b) Atom splitting probability between tweezers B and C with trap depth ratio. The dotted lines indicate where 33~$\%$ of atoms are split into tweezer C. (c) Examples of inelastic loss features. Data are post-selected to show only experimental runs where an atom pair was initially present. (i) Probability of detecting zero atoms after trap separation  (ii) Probability of detecting one atom. (iii) Probability of detecting two atoms.}
	\label{fig:collisions:FR_measurement_paper}
\end{figure}

\noindent
The apparatus used for the experiments reported here has been discussed in refs.~\cite{Brooks2021, Spence2022}. In this section we revisit the relevant aspects of the apparatus and detail new elements that are instrumental in performing the measurements presented here. Figure~\ref{fig:collisions:FR_measurement_paper}(a) shows a simplified overview of the experimental sequence we use to probe two-body collisions.  We first form a partially filled array of 938~nm optical tweezers and rearrange it to produce a pair of tweezers A and B with a single Cs atom in each. Both Cs atoms are then transferred to a 1064~nm ``collision tweezer''. Following a hold time in the collision tweezer at a magnetic field $B$, the remaining atoms are transferred back to a 938~nm tweezer, which is then sub-divided into 3 separate tweezers. The occupation of these ``imaging tweezers'' is detected using fluorescence imaging.

\subsection{Preparation of Cs Atom Pairs}
\label{subsec:prep}

To prepare pairs of Cs atoms, we first load atoms from a magneto-optical trap (MOT) into a 1D array of 5 optical tweezers of wavelength 938~nm. The probability that an atom is loaded into any one tweezer from the red-detuned MOT is only  about 50\% \cite{Schlosser2002}, so the likelihood of preparing at least two atoms is significantly enhanced by loading multiple tweezers. The tweezer array is produced using an acousto-optic deflector (AOD) driven by the sum of 5 radio-frequency (RF) sine waves, each with a distinct frequency and phase. The composite wave is produced digitally by a software-controlled arbitrary waveform generator (AWG). The multiple diffracted beams are focussed by a high-numerical-aperture objective to form a tweezer array centred on the Cs MOT. The tweezer spacing is proportional to the frequency difference between adjacent tones, which can be tuned to zero. The bandwidth of the AOD and magnification of the imaging system allows a maximal array extent of $29~\mu$m. The spacing between adjacent tweezers in the 5-tweezer array is 4~$\mu$m, corresponding to a frequency separation of 12.5~MHz between the RF tones. For our parameters, the AOD diffraction efficiency per tweezer is $\sim 8~\%$. Using the maximum laser power available, we produce tweezers of depth $U/k_\mathrm{B}=0.26$~mK, which is sufficient to saturate the loading probability of each tweezer; we obtain a mean probability of 0.53(1) across the array \cite{Brooks2021}. 

We rearrange the 5-trap array to prepare a 2-atom array. First, the occupancy of the 1D array is determined by fluorescence imaging. Then, the rearrangement procedure is performed by extinguishing RF tones corresponding to empty tweezers and those containing excess atoms, before translating the remaining occupied tweezers to their target positions \cite{Endres2016}. The tweezers are shuttled in 1D by dynamically tuning the RF frequency of the corresponding tones simultaneously in 1~ms, following a minimum-jerk trajectory to reduce heating. After rearrangement, the mean atom temperature in each tweezer is $7(1)~\mu$K. We observe that each experimental shot has a probability of 0.84(2) of preparing a 2-atom array. This agrees with the expected binomial probability calculated from the mean single-site loading probability of the array; it corresponds to near-unity probability of preparing 2 atoms in the specified tweezers whenever there are 2 or more atoms in the 5-trap array. For the measurements of Feshbach resonances, experimental shots are post-selected on events where exactly two Cs atoms are prepared following the rearrangement.

Both atoms are optically pumped (OP) to one of the target states $(f=3, m_f=\pm3)$ using an OP beam aligned along the $x$ axis. The probability of preparing each atom in the target state is 0.99(1), resulting in a joint probability of preparing both atoms in the target state of 0.98(2).

\subsection{Pair Transfer to a Collision Tweezer}
\label{subsec:transfer}
The individual Cs atoms are transferred from their initial 938~nm tweezers, labelled A and B, to a single ``collision tweezer'' of wavelength 1064~nm (Fig.~\ref{fig:collisions:FR_measurement_paper}(a)(iii)). This wavelength avoids spontaneous Raman scattering of tweezer photons, which otherwise causes depopulation of the target state. For the same trap depth, the spontaneous Raman scattering rate in a 1064~nm tweezer is a factor of about 100 less than for a 938~nm tweezer due to the greater detuning from the Cs D$_1$ and D$_2$ transitions. Furthermore, in contrast to merging two 938~nm tweezers using the AOD, parametric heating effects from the beating between adjacent RF tones \cite{Endres2016} are removed by merging into a collision tweezer of a distinctly different wavelength. The collision tweezer overlaps with tweezer B. 
The position of the collision tweezer is controlled in 3 dimensions by modifying the phase profile of the 1064~nm beam using a spatial light modulator (SLM). The overlap of the collision tweezer with the 938~nm tweezer is optimised by measuring atom transfer probabilities between the tweezers, as in our previous work \cite{Brooks2021}.

Once preparation of the atom pair is complete, the collision tweezer is ramped up to a depth of $U_\mathrm{coll}/k_\mathrm{B}=0.35$~mK in 5~ms. Tweezer B is then adiabatically ramped off in $3$~ms, transferring its atom to the collision tweezer. Next, the frequency of tweezer A is swept to overlap it with the collision tweezer in 2~ms, where it too is ramped off in 3~ms. Following the merging process, only the 1064~nm collision tweezer containing a pair of Cs atoms remains.

The transfer to the collision tweezer must occur with minimal heating in order to maintain a high pair density, which decreases with atom temperature as $n_2 \propto T^{-3/2}$. We use a hybrid jerk trajectory \cite{Liu2019} to merge the tweezers. Its profile is designed to minimise heating from sudden acceleration of the tweezers and from sweeping at resonant frequencies of the AOD \cite{Savard1997, Spence2022}. The tweezer depth, $U_{\rm A}$,  must also be set to be equal to the depth of the collision tweezer, $U_\mathrm{coll}$, to avoid spilling during merging. If an atom spills into a deeper tweezer, it gains kinetic energy along the merge axis roughly equal to the difference in trap depths. To equalise the tweezer depths we measure the kinetic energy of the atoms after they are transferred to the collision tweezer using a release-and-recapture technique described elsewhere \cite{Tuchendler2008, Brooks2021}. $U_{\rm A}$ is then equalised with $U_\mathrm{coll}$ by adjusting the amplitude of the corresponding RF tone applied to the AOD until the energy gained during merging by an atom prepared in either tweezer can no longer be resolved.

Once the atom transfer is complete, the collision tweezer is adiabatically lowered to a depth of $30~\mu$K in 5~ms for the duration of the variable collision time $t_\mathrm{coll}$. The harmonic-oscillator energy $E_\mathrm{HO}$ in a tweezer scales as $\sqrt{U}$, so the decrease in tweezer depth reduces the trap frequencies to $\{ \nu_x, \nu_y, \nu_z \} = \{7.4, 9.1, 1.3\}$~kHz (estimated by scaling the trap frequencies measured in a deeper trap). The atomic temperature also reduces to $1.7(3)~\mu$K. The pair density, given by $n_2 = \int n_\mathrm{A}(r) n_\mathrm{B}(r) dr^3$, quantifies the spatial overlap of the atoms transferred from A and B into the collision tweezer. For the trap frequencies and temperature noted above, and assuming that the energies of the individual atoms averaged over many iterations of the experiment obey a Boltzmann distribution, we obtain a pair density $n_2=5(2) \times 10 ^{11}$~cm$^{-3}$. 

In the following, we probe the two-body collisions at a variable magnetic field $B$ aligned along the merge axis $x$ (Fig.~\ref{fig:collisions:FR_measurement_paper}(a)(iv)); this is ramped up in 10~ms during the tweezer merging step and is maintained for the duration of the collision time. 

\subsection{Imaging Homonuclear Atom Pairs}
\label{subsec:imaging}

The outcome of any collision process that results in the loss of one ($2 \to 1$ loss) or both ($2 \to 0$ loss) atoms from the collision tweezer is detected by splitting the collision tweezer into multiple ``imaging tweezers'' whose occupancy is probed using a final pulse of resonant imaging light. We first transfer the atoms that remain in the collision tweezer back to tweezer B. The AOD is then used to subdivide tweezer B repeatedly by splitting off a second tweezer and sweeping its position. The probability that an atom remains in tweezer B and is not transferred to the new tweezer, $P_\mathrm{split}$, is set by the relative depth of the two tweezers as shown in Fig.~\ref{fig:collisions:FR_measurement_paper}(b). The splitting process is repeated multiple times to produce a 1D array of $N$ imaging tweezers. The results of the imaging procedure are binned according to whether 0, 1 or 2 atoms are detected. Two atoms are detected only when they are trapped in separate imaging tweezers. In cases where two atoms are initially in the same imaging tweezer, rapid light-assisted collisions induce either $2 \to 0 $ loss with probability $P^{\rm{img}}_{2 \to 0}$ or $2 \to 1$ loss with probability ${P^{\rm{img}}_{2 \to 1} =1-P^{\rm{img}}_\mathrm{2\to0}}$, contributing to the observed background in these detection channels. We refer to this as ``imaging loss''. We measure the probability of these events occurring for our experimental parameters by preparing two atoms in the same imaging tweezer. We measure $P^{\rm{img}}_{2 \to 0} = 0.64(2)$ and $P^{\rm{img}}_{2 \to 1} =0.36(2)$. In other experiments it is possible to observe multiple particles in the same optical tweezer \cite{McGovern2011, Anderegg2019, Jackson2020}, but the imaging losses preclude such an approach in our experiment.

The detection scheme used here is improved by increasing the number of division steps, because the probability for both atoms to occupy the same final tweezer decreases as $N$ increases. 
Splitting is currently limited to $N=3$, as indicated in Fig.~\ref{fig:collisions:FR_measurement_paper}(a)(v). This is because the splitting probability depends on the temperature-dependent diffraction efficiency of the AOD. As the experimental sequence cycles between 5, 2, 0 and 3 traps, the amount of RF power dissipated in the AOD changes. When $t_\mathrm{coll}$ is varied, the mean power dissipated during a sequence changes; this affects the splitting probabilities and requires rebalancing of the depths. This becomes inconvenient as $N$ is increased. An upgrade of the tweezer laser system to increase the available power would allow the use of lower RF drive power for the AOD, simplifying the balancing process and improving the detection scheme.

\subsection{Detection of Feshbach Resonances}
\label{subsec:detection}

The inelastic two-body collision rate is field-dependent and increases sharply in the vicinity of a Feshbach resonance. Feshbach resonances may therefore be detected as an enhancement in the $2 \to 0 $ loss rate. We perform Feshbach spectroscopy by probing the pair loss as a function of magnetic field. The one-body $1/e$ lifetime is greater than 20~s, which is much longer than  $t_\mathrm{coll}\leq 250$~ms, and so is not expected to influence these measurements.

Loss signals typical of our detection scheme  are shown in Fig.~\ref{fig:collisions:FR_measurement_paper}(c). The results of the imaging procedure are
binned according to whether 0, 1 or 2 atoms are detected. The probabilities $P_0$, $P_1$ or $P_2$ of 0, 1 or 2 atoms surviving are shown in Fig.~\ref{fig:collisions:FR_measurement_paper}(c)(i-iii) respectively. Here, the solid lines are Lorentzian fits to the data to determine the centres of the features. The enhancement in the collisional loss rate near the Feshbach resonance is observed as a minimum in $P_2$ and a peak in $P_0$. The background level of the feature in $P_2$ extracted from the fit is $0.38(2)$, which corresponds to the probability that the atom pair survived the hold time and was split into separate tweezers for imaging. We also observe a reduction in probability of detecting one atom, $P_1$, near the loss feature. As described above, $P_1$ corresponds to events where both atoms remain in the collision tweezer after $t_\mathrm{coll}$ but are subsequently mapped onto $P_1$ by $2 \to 1$ loss during the imaging stage. Increased collisional loss of both atoms during $t_\mathrm{coll}$ therefore leads to a correlated reduction in $P_1$ below the background value for fields near the loss feature.

In the remainder of our study of Feshbach spectroscopy, we use the $P_0$ feature unless explicitly stated otherwise. This avoids complications from (unlikely) $2 \to 1$ collisional loss and the 1-body $1/e$ lifetime.

\section{Results: Feshbach Spectroscopy} 
\label{sec:results}

In this section, we present the results of Feshbach spectroscopy of Cs atom pairs in an optical tweezer.  We observe 5 loss features for Cs pairs prepared in $(f=3,m_f=-3)$, using the inelastic loss spectroscopy described above. In addition, we observe 4 features for Cs pairs prepared in $(f=3,m_f=+3)$, using radiative loss spectroscopy, which we describe in section~\ref{subsec:radiative}. These features have been characterised in previous works \cite{Vuletic1999, Chin2004, Mark2018}, and so serve as a benchmark for our tweezer-based measurement scheme.

We follow the notation of ref.~\cite{Berninger2013} and label each feature by the quantum numbers $n(f_1f_2)FL(M_F)$ of the bound state responsible for the Feshbach resonance. Here $(f_1 f_2)$ indicates the atomic threshold that supports the bound state, $F$ is the resultant of $f_1$ and $f_2$, $M_F=m_{f,1}+m_{f,2}$ and $L$ is the rotational angular momentum of the bound state, labelled  with $\{\mathrm{s, d, g,...}\}$ to indicate $L=\{0,2,4,...\}$; the molecular vibrational quantum number $n$ is counted with respect to the threshold $(f_1 f_2)$, with $n=-1$ being the least-bound state.

\subsection{Inelastic Feshbach Spectroscopy of $(f=3,m_f=-3)$ Pairs}\label{subsec:3-3}

\begin{figure}
	\centering
	\includegraphics[width=0.9\linewidth]{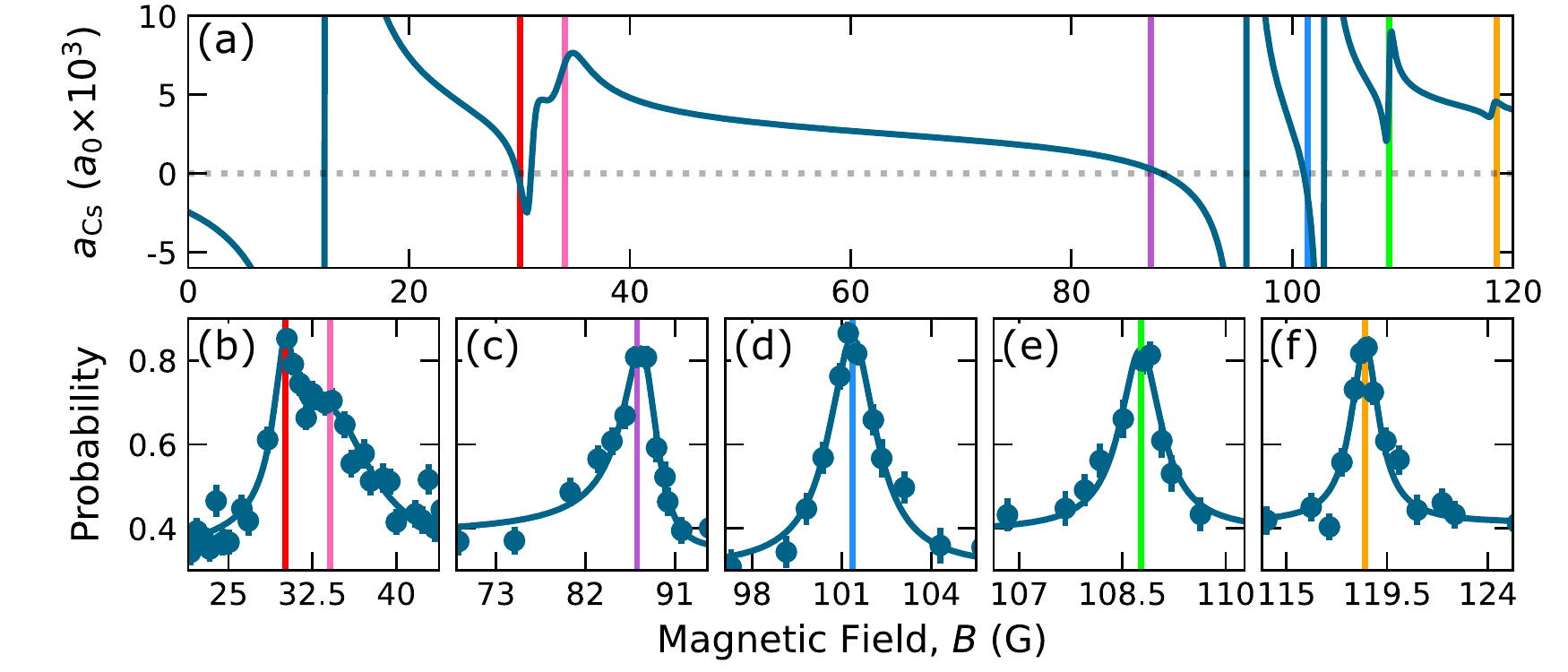}
	\caption[Feshbach spectroscopy using Cs atom pairs prepared in $(f=3,m_f-3)$]{Inelastic Feshbach spectroscopy of Cs atom pairs prepared in \mbox{$(f=3,m_f=-3)$}. (a) Real part of the calculated s-wave scattering length as a function of applied magnetic field \cite{Frye2019}. The coloured lines correspond to the features measured in (b)-(f). (b)-(f) Probability of observing zero atoms after a hold time at field $B$. (b) Loss features around 30 and 34~G for $t_\mathrm{coll}=50$~ms. (c) Loss feature around 88~G for $t_\mathrm{coll}=250$\,ms. (d) Loss feature around 101~G for $t_\mathrm{coll}=50$~ms. (e) Loss feature around 109~G for $t_\mathrm{coll}=50$~ms. (f) Loss feature around 119~G for $t_\mathrm{coll}=250$~ms.}
	\label{fig:collisions:3-3FR}
\end{figure}

The experimental results of the Feshbach spectroscopy for atom pairs prepared in $(f=3,m_f=-3)$ are shown in Fig.~\ref{fig:collisions:3-3FR}. For reference, panel (a) shows the real part of the s-wave scattering length as a function of magnetic field; this is taken from the coupled-channel calculations of ref.~\cite{Frye2019}, based on the most recent interaction potential for Cs \cite{Berninger2013}. The vertical lines mark the fitted centres of the loss features shown in Fig.~\ref{fig:collisions:3-3FR}(b-f), which plot $P_0$ as a function of magnetic field in the vicinity of each feature. For these measurements the collision time was adjusted to give $P_0\simeq0.8$ at the peak of each feature. We note that the loss feature in Fig.\ \ref{fig:collisions:3-3FR}(c) is significantly weaker than those for the other resonances. To observe this feature we used a longer hold time of 250~ms and a tweezer depth of $U_\mathrm{coll}/k_\mathrm{B}=0.33$~mK.  

Thermal broadening is expected to contribute only about 50~mG to the linewidth of the loss features shown in Fig.~\ref{fig:collisions:3-3FR}(b-f). This is much narrower than the observed widths. We therefore use a Lorentzian fit to extract the centre $B_0$ and the full width at half maximum $\Gamma$ of the features, yielding reduced $\chi^2$ fitting parameters close to 1 for all except the asymmetric feature in Fig.\ \ref{fig:collisions:3-3FR}(c); this is conveniently fitted with a generalised Fano profile.

The doubly peaked feature in Fig.~\ref{fig:collisions:3-3FR}(b) corresponds to two closely spaced resonances with fitted centres at 30.1(2)~G and 34.1(8)~G, caused by the threshold crossing of the $6 \mathrm{d} (-6)$ and $6 \mathrm{d} (-4)$ bound states respectively \cite{Chin2004}. For this pair, an abbreviated notation $FL(M_f)$ is used to describe the relevant bound states because the $n(f_1 f_2)$ quantum numbers are not well  defined. For the feature in Fig.~\ref{fig:collisions:3-3FR}(c) we extract a centre of 88.0(3)~G, in agreement with previous measurements in bulk gases \cite{Leo2000, Chin2004}. The features shown in Fig.~\ref{fig:collisions:3-3FR}(d-f) arise from the threshold crossings of the d-wave bound states $-7(44)8 \mathrm{d} (-7)$, $-7(44)8 \mathrm{d} (-6)$ and  $-7(44)8 \mathrm{d} (-5)$ respectively. The widths and centres extracted from fits to the observed features are summarised in table~\ref{table:FRs}. The fitted positions are in agreement with the values obtained in ref.~\cite{Chin2004}. To our knowledge, the extracted widths have not been listed previously.

\subsection{Feshbach Spectroscopy of $(f=3,m_f=+3)$ Pairs}
\label{subsec:radiative}

\begin{figure}
	\centering
	\includegraphics[width=0.9\linewidth]{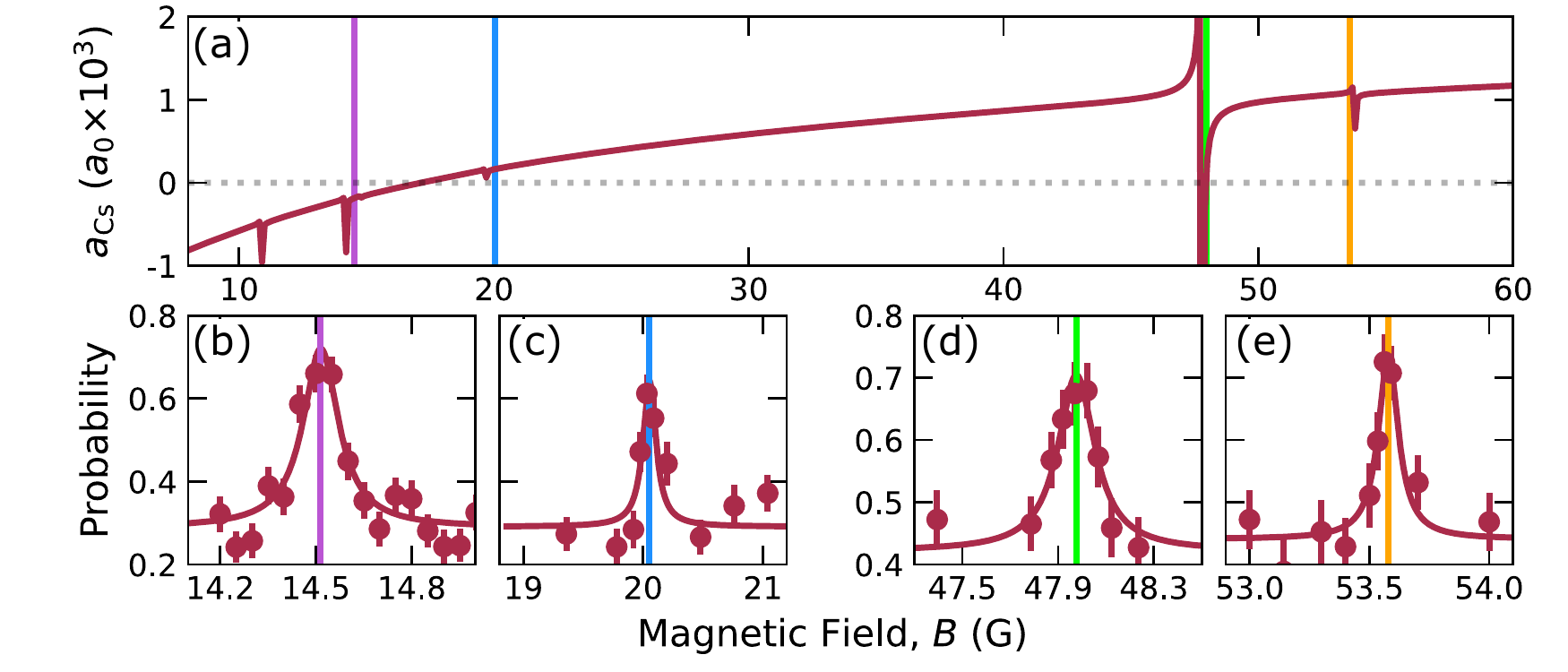}
	\caption[Radiative Feshbach spectroscopy of Cs atom pairs prepared in $(f=3,m_f=3)$.]{Radiative Feshbach spectroscopy of atom pairs prepared in the state $(f=3,m_f=3)$. (a) Calculated s-wave scattering length as a function of applied magnetic field \cite{Frye2019}; narrow resonances are not fully resolved due to the field grid of 0.1~G. The coloured lines correspond to the loss features shown  in (b)-(e).  (b) A beam blue-detuned by 47~GHz induces loss of molecules during a hold time at field $B$. The probability of observing zero atoms is plotted. Loss feature around 14.4~G. (c) Loss feature around 20~G. (d) Loss feature around 48~G. (e) Loss feature around 53.5~G. }
	\label{fig:collisions:33FR}
\end{figure}

Two-body inelastic collisions cannot occur when the atoms are in their lowest internal state $(f=3,m_f=3)$. We therefore use radiative loss spectroscopy \cite{Vuletic1999, Chin2003} to probe the Feshbach spectrum. We apply an excitation beam, blue-detuned by $\delta_\mathrm{ex}$ from the Cs D$_2$ line, to couple the atom pair in the ground state to an electronically excited molecular state with a repulsive interaction.
The excess energy $h\delta_\mathrm{ex}$ is converted into relative kinetic energy of the atoms and ejects them from the trap.

The radiative loss is strongly modified in the vicinity of a Feshbach resonance. The repulsive interaction in the excited state is of the form $C_3/R^3$, where $C_3$ is the coefficient of the resonant electric dipole-dipole interaction \cite{Jones2006}, calculated in ref.~\cite{Fioretti1999}. Excitation occurs around the Condon radius $R_\mathrm{C} \approx [C_3/(h \delta_\mathrm{ex})]^{1/3}$ \cite{Burnett1996}. We use an excitation beam with $\delta_\mathrm{ex} =47.2$~GHz, which is applied for the duration of the hold in the collision tweezer. The energy released is many orders of magnitude larger than the depth of the collision tweezer, so that pairwise loss is guaranteed if excitation occurs. For Cs at this detuning, $R_\mathrm{C} \approx 60$~\AA, and the probability of finding a free atom pair at such short range is very low. However, in the vicinity of a Feshbach resonance, the scattering state acquires some character of the bound state responsible for the resonance. This enhances the amplitude of the scattering wavefunction at small distances and thus enhances the probability of optical excitation and loss.

Radiative loss Feshbach spectroscopy for Cs pairs prepared in the $(f=3,m_f=+3)$ state is shown in Figure.~\ref{fig:collisions:33FR}. Panel (a) shows the s-wave scattering length of this state as a function of magnetic field \cite{Frye2019}. The coloured lines indicate the fitted positions of the loss features shown in panels (b-e) and summarised in table~\ref{table:FRs}.

\begin{table}
\caption{Fitted centres of loss features observed for Cs in the states $(f=3,m_f=\pm3)$. The centres and widths extracted from fits as described in the text are given in the columns headed $B_0^\mathrm{expt}$ and $\Gamma^\mathrm{expt}$ respectively. \label{table:FRs}} 
\begin{indented}
\item[]\begin{tabular}{@{}crrr}
\br
$(f,m_f)$ & $n(f_1f_2)FL(M_f)$ & $B_0^\mathrm{expt}$~(G) & $\Gamma^\mathrm{expt}$~(G)\\
\mr
$3,-3$ &  $6 \mathrm{d} (-6)$  &   30.1(2)  &  2.4(8) \\
     & $6 \mathrm{d} (-4)$     &  34.1(8)   &  9(3) \\
     &  -             &       88.0(3)   &    4.6(3)    \\
     &  $-7(44)8 \mathrm{d} (-7)$  &  101.4(1)   &  1.9(2) \\
     & $-7(44)8 \mathrm{d} (-6)$ &    108.77(6)  &  0.7(1)\\
     & $-7(44)8 \mathrm{d} (-5)$ & 118.51(5)  &  1.5(2)\vspace{3pt}\\
\hline

$3,+3$ &  $-2 (33)4\mathrm{g} (3) $ &  14.52(5)  & 0.14(3)\\
     & $-2 (33)4\mathrm{g} (4)  $ &  20.02(2)   &  0.13(6) \\
    & $-2 (33)4 \mathrm{d} (4)$  & 47.98(2)   & 0.19(5) \\
    & 	$x2\mathrm{g}(2)   $  &  53.58(2)   &   0.10(3)\\

\br
\end{tabular}
\end{indented}
\end{table}

The elastic widths $\Delta$ of the resonances at 14.5~G and 20.0~G have previously been measured and are small ($\Delta \lesssim 15$~mG) \cite{Mark2018} because of the weakness of the second-order spin-orbit coupling. 
However, in the present case the excitation laser provides a loss channel with its own inelastic width, which may be larger than the elastic width, effectively converting the elastic resonances into decayed resonances. The elastic and inelastic widths interact to produce complicated loss profiles, but the inelastic parameters involved are hard to quantify theoretically. The profiles are also modified by thermal effects. We therefore fit each loss feature using a Lorentzian function to extract its centre $B_0^\mathrm{expt}$ and experimental width $\Gamma^\mathrm{expt}$.

Our measured peak positions are in good agreement with those observed in bulk gases \cite{Chin2004} and lattices \cite{Mark2018}. However, as may be seen in Fig.~\ref{fig:collisions:33FR}(a), there are small but significant shifts from the theoretical zero-energy resonance positions obtained using the interaction potential of Berninger \emph{et al.} \cite{Berninger2013}. There may be some contribution to these differences from thermal and radiative shifts, but they are comparable to some of those found for other low-field features in ref.\ \cite{Berninger2013}, so may also be due to deficiencies in the interaction potential.

\section{Theory and Discussion}\label{sec:theory}

In order to understand the loss features for the $(3,-3)$ state, we have carried out coupled-channel scattering calculations on the interaction potential of Berninger \emph{et al.}\ \cite{Berninger2013}, using the \textsc{molscat} package \cite{molscat:2019, mbf-github:2020}. The methods used are similar to those in Ref.\ \cite{Berninger2013}, so only a brief outline is given here.

The Hamiltonian for the interacting pair is
\begin{equation}
\label{full_H}
\hat{H} =\frac{\hbar^2}{2\mu}\left[-\frac{1}{R}\frac{d^2}{dR^2}R
+\frac{\hat{L}^2}{R^2}\right]+\hat{H}_\textrm{A}+\hat{H}_\textrm{B}+\hat{V}(R),
\end{equation}
where $R$ is the internuclear distance, $\mu$ is the reduced mass, and $\hbar$ is the reduced Planck constant. $\hat{L}$ is the two-atom rotational angular momentum operator. The single-atom Hamiltonians $\hat{H}_i$ contain the hyperfine couplings and the Zeeman interaction with the magnetic field $B$.
The interaction operator $\hat{V}(R)$ contains the two isotropic Born-Oppenheimer potentials, for the X $^1\Sigma_g^+$ singlet and $a$ $^3\Sigma_u^+$ triplet states, and anisotropic spin-dependent couplings which arise from magnetic dipole-dipole and second-order spin-orbit coupling. The scattering wavefunction is expanded in a fully uncoupled basis set that contains all allowed spin functions, limited by $L_{\rm max}=4$. Solutions are propagated from $R_\textrm{min}=6\, a_0$ to $R_\textrm{mid}=20\, a_0$ using the diabatic modified log-derivative propagator of Manolopoulos \cite{Manolopoulos:1986} with a step size of 0.002 $a_0$, and from $R_\textrm{mid}$ to $R_\textrm{max}=10000\, a_0$ using the log-derivative Airy propagator of Alexander and Manolopoulous \cite{Alexander:1987} with a variable step size. The log-derivative matrix is transformed into the asymptotic basis set at $R_\textrm{max}$ and matched to S-matrix boundary conditions to obtain the scattering matrix ${\bf S}(k,B)$, where $k=\sqrt{2\mu E_\textrm{coll}}/\hbar$ is the incoming wavevector and $E_\textrm{coll}$ is the collision energy. 

\begin{figure}
	\centering
	\includegraphics[width=0.9\linewidth]{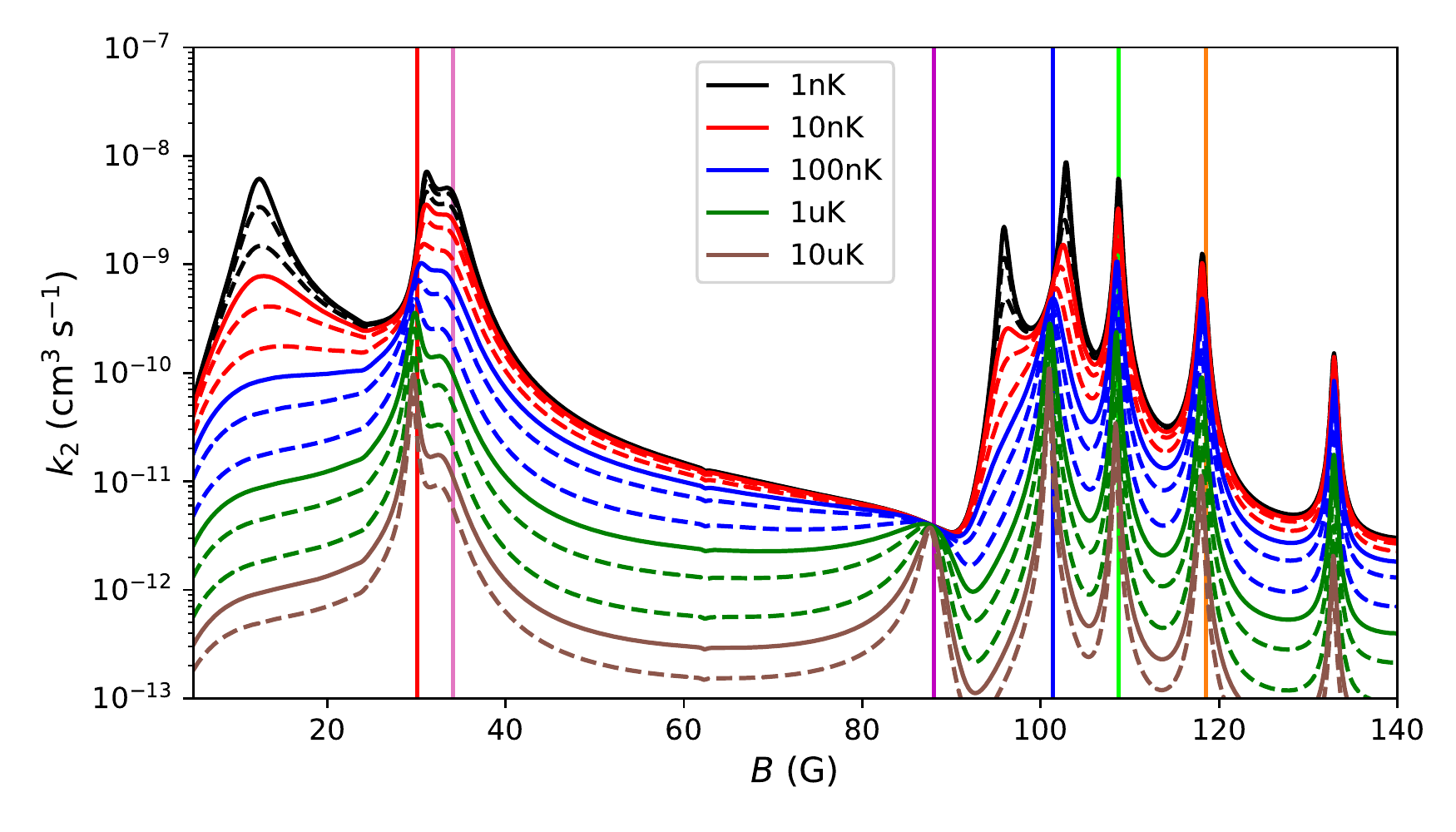}
	\caption{Overview of results of coupled-channel scattering calculations on Cs atom pairs in $(f=3,m_f=-3)$. The two-body loss rate coefficient at $E/k_\textrm{B}=1$ \textrm{nK} (black), 10 nK (red), 100 nK (blue), 1 $\mu$K (green), and 10 $\mu$K (brown); dashed lines are 2 and 5 times the energy of the corresponding solid lines.}
	\label{fig:theory_overview}
\end{figure}

\begin{figure}
	\centering
	\includegraphics[width=0.9\linewidth]{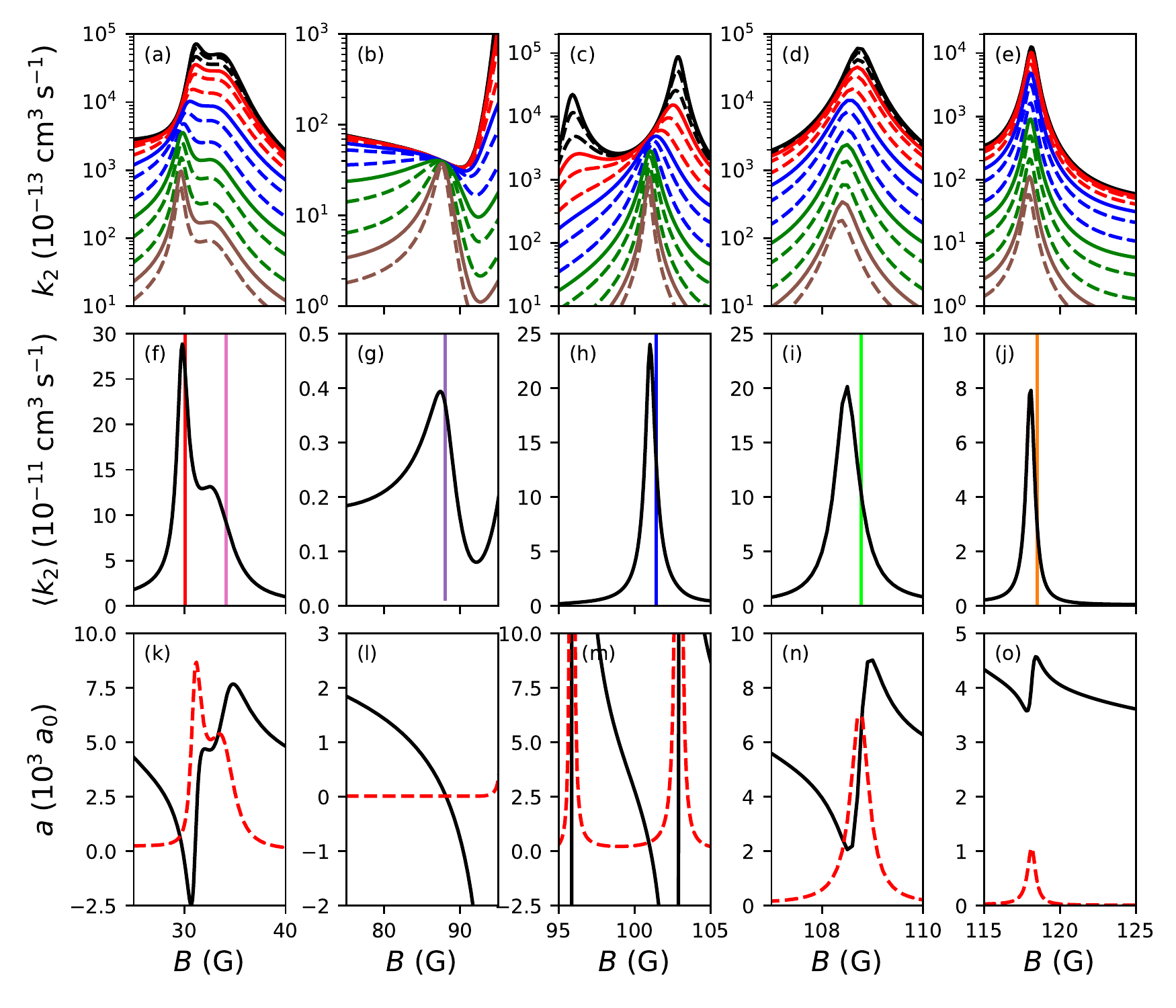}
	\caption{Results of coupled-channel scattering calculations on Cs atom pairs in $\ket{f=3,m_f=-3}$. (Top row, a-e) Expanded views of segments of figure \ref{fig:theory_overview}, corresponding to the experimental results in Fig.~\ref{fig:collisions:3-3FR}, with the same line types and colours for rate coefficients. (Middle row, f-j) Thermally averaged loss-rate coefficient at $T=1.7\ \mu\textrm{K}$; note the linear scale. The vertical coloured lines match those in Fig.\,\ref{fig:collisions:3-3FR} and indicate the measured peak positions. (Bottom row, k-o) Complex scattering length $a=\alpha-\textrm{i}\beta$ at limitingly low energy, showing $\alpha$ (solid black lines) and $\beta$ (dashed red lines).
	\label{fig:theory}
	}
\end{figure}

We have carried out calculations of the rate coefficient $k_2$ for 2-body inelastic loss from the $(f=3,m_f=-3)$ state to all lower-energy states. The calculations are carried out on approximately logarithmic grid of collisions energies from $1~\textrm{nK}\times k_\textrm{B}$ to $10~\mu\textrm{K}\times k_\textrm{B}$. Only contributions from the incoming s-wave are included; the height of the d-wave centrifugal barrier is $180~\mu\textrm{K} \times k_\textrm{B}$, and d-wave contributions to $k_2$ are generally small at the collision energies considered here.
The s-wave contribution to $k_2$ is \cite{Hutson:res:2007}
\begin{equation}
\label{eq:k2}
k_2 = \frac{4 \pi \hbar \beta}{\mu (1 + k^2 \vert a \vert ^2 + 2 k \beta)},
\end{equation}
where $\mu$ is the reduced mass and $a(k,B)= \alpha(k,B) - i \beta(k,B)$ is the complex energy-dependent scattering length. This is obtained from the diagonal S-matrix element $S_{00}(k,B)$ for the incoming channel,
\begin{equation}
a(k,B) = \frac{1}{ik} \left(\frac{1-S_{00}(k,B)}{1+S_{00}(k,B)}\right).
\end{equation}
Both $\alpha(k,B)$ and $\beta(k,B)$ have significant energy dependence, which is taken into account in the calculations, but it has only weak effects on $k_2$, so in the following we refer to $\alpha(B)$ and $\beta(B)$ for simplicity.
For pure elastic scattering, the scattering length is real and resonances appear as poles in $a(B)$ as a function of magnetic field. However, when inelastic scattering can occur, $a(B)$ remains finite and each resonance is characterized by a resonant scattering length $a_\textrm{res}$ \cite{Hutson:res:2007}, which can be efficiently extracted from our calculations \cite{Frye:resonance:2017}. When the background inelastic scattering is fairly weak, as here, $\beta(B)$ typically shows a peak of height $a_\textrm{res}$ near resonance, and $\alpha(B)$ shows an oscillation of amplitude $\pm a_\textrm{res}/2$, which may or may not cross zero.

Figure \ref{fig:theory_overview} shows an overview of the loss rate coefficients for the $(f=3,m_f=-3)$ state over the full range of magnetic fields considered here, while Figs.~\ref{fig:theory}(a) to (e) show expanded regions corresponding to the experimental results in Fig.~\ref{fig:collisions:3-3FR}. Figs.~\ref{fig:theory}(f) to (j) show the thermally averaged loss rate $\langle k_2 \rangle = \int (2/\pi) k_2(E_\textrm{coll}) x^{1/2} e^{-x} dx$, where $x=E_\textrm{coll}/(k_\textrm{B}T)$.
Figs.~\ref{fig:theory}(k) to (o) show $\alpha(B)$ and $\beta(B)$, evaluated at limitingly low collision energy. At the lowest energies, the term in parentheses in the denominator of Eq.\ \ref{eq:k2} is close to 1 and the loss rate shows a peak centred at the resonance position. However, all states of Cs have background scattering lengths (far from resonance) that are large and positive. Because of this, the denominator of Eq.\ \ref{eq:k2} causes significant suppression of the loss rate at most fields for collision energies above $100~\textrm{nK}\times k_\textrm{B}$.
Even for the moderately low collision energies in the present experiment ($E_\textrm{coll}/k_\mathrm{B} \approx 2~\mu$K), the loss peaks mostly appear near the zeroes in $\alpha(B)$, rather than at the resonance centres where there are peaks in $\beta(B)$.

This effect is most evident for the loss peak near 88.0~G, which is far from any resonance and is centred around a zero in $\alpha(B)$; the development of the peak as a function of $E_\textrm{coll}$ is shown in Fig.~\ref{fig:theory}(b). However, even for the narrower resonance near 101.4~G, shown in Fig.~\ref{fig:theory}(c), the loss peak clearly shifts from its zero-energy field as $E_\textrm{coll}$ increases, and is close to the zero in $\alpha(B)$ at the experimental energy. A similar shift occurs for the resonance near 108.77~G, shown in Fig.~\ref{fig:theory}(d). This resonance is more strongly decayed, with $a_\textrm{res}\sim 7100~a_0$, and $\alpha(B)$ does not actually cross zero; nevertheless, it is the minimum in $\alpha(B)$ that determines the peak position, rather than the peak in $\beta(B)$. The loss features between 30 and 35~G, shown in Fig.~\ref{fig:theory}(a), are more complicated; the main experimental peak at 30.1~G occurs near the zero in $\alpha(B)$, while the shoulder near 34.1~G is principally due to a peak in $\beta(B)$, in a region where $\alpha(B)$ does not have a pronounced minimum. The peak at 118.51~G, shown in Fig.~\ref{fig:theory}(e) is an exception to the general rule; the corresponding resonance is strongly decayed ($a_\textrm{res}\sim 1000~a_0$), so there is little variation in the denominator of Eq.\ \ref{eq:k2} and the loss peak is due to the peak in $\beta(B)$.
The resonances near 15~G and 95~G are clearly visible at 1~nK in figure \ref{fig:theory_overview}, but are suppressed by this effect at 1~$\mu$K and above. This agrees with our experimental measurements, which show no evidence of these resonances at our experimental temperature.

Panels (f)-(j) of Fig.\ \ref{fig:theory} show the measured peak positions as coloured lines, corresponding to those in Fig.\ \ref{fig:collisions:3-3FR}. In this case the theoretical profiles take full account of thermal broadening, so the small remaining difference between the experimental and theoretical peak positions may be attributed to deficiencies in the interaction potential of ref.\ \cite{Berninger2013}.

\section{Probing the Energy Release of $(f=3,m_f=-3)$ Collisions}\label{sec:depth}

Another application of our experimental platform is to investigate the kinetic energy released in inelastic collisions. To do this, we follow the same general experimental scheme as described in section \ref{sec:experimental_methods}. However, we change the depth of the collision tweezer $U_\textrm{coll}$, such that the atoms may remain trapped if they do not gain enough kinetic energy to escape. We can thus probe this kinetic energy release by observing $P_0$, $P_1$, and $P_2$ while varying $U_\textrm{coll}$, and comparing them with coupled-channel calculations.

The kinetic energy released from an inelastic collision is shared equally between the two atoms in the centre-of-mass frame. The initial kinetic energies are much smaller than the energy release, so the resulting lab-frame energies are almost equal.
The states formed by the inelastic collisions are dominated by those with $\Delta M_F=+1$ and $+2$ because these can be accessed from the incoming state by couplings first-order in the anisotropic spin-dependent couplings.
For a collision with $\Delta M_F=+1$, the energy each atom receives is $\Delta E_{1}=\frac{1}{2}\mu_{\rm B}g_{\rm F}B$.
If $U_\textrm{coll} < \Delta E_1$ such a collision leads to both atoms leaving the tweezer rapidly; if $\Delta E_1 < U_\textrm{coll} < 2\Delta E_1$, the atoms remain trapped, at least initially, even though there is sufficient total kinetic energy for one atom to escape; if $2\Delta E_1 < U_\textrm{coll}$ then a single inelastic collision should not cause loss. For a collision with $\Delta M_F=+2$, the kinetic energy release is doubled.\footnote{For Cs at low fields, the quadratic Zeeman effect is negligible, so this value does not depend on the states of the individual atoms.}

\begin{figure}
	\centering
	\includegraphics[width=0.9\linewidth]{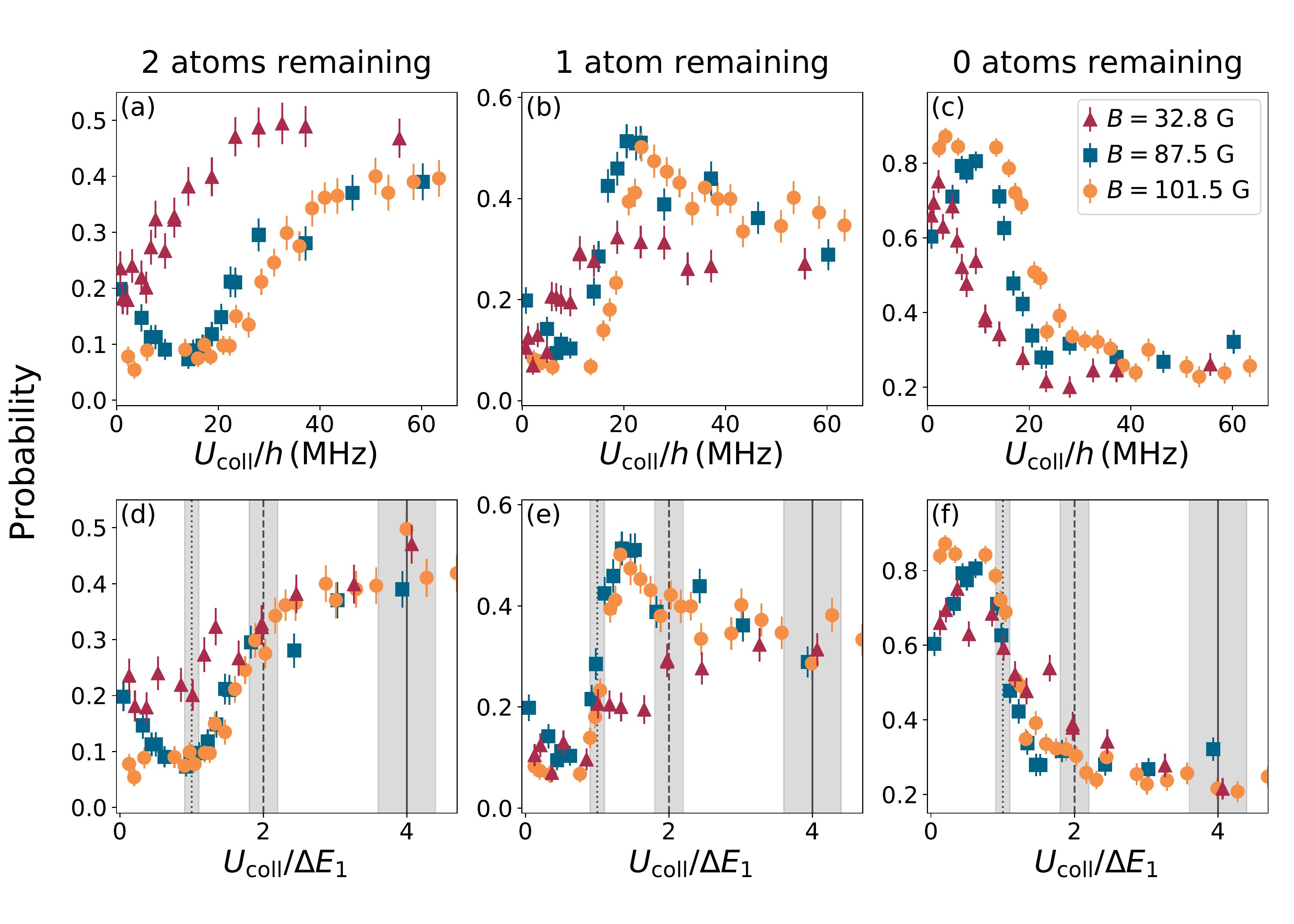}
	\caption[Effect of trap depth on atom loss from the collision tweezer]{Effect of the collision tweezer depth, $U_\mathrm{coll}$, on atom loss at three different magnetic fields. The left, centre and right columns show the probability of observing 2 atoms, 1 atom and 0 atoms respectively. Measurements at 32.8~G, 87.5~G and 101.5~G are shown as red triangles, blue squares and orange circles, respectively. (a)-(c) show results as a function of $U_\mathrm{coll}/h$ in units of MHz. (d)-(f) show the same results with the x axis rescaled by $\Delta E_{1}=\frac{1}{2}\mu_{\rm B}g_{\rm F}B$. The vertical lines indicate different multiples of $\Delta E_{1}$, while the shaded regions represent the 1$\sigma$ uncertainty on the absolute value of $U_\mathrm{coll}$ at these tweezer intensities.}
	\label{fig:collisions:Trap_depth}
\end{figure}

Figure~\ref{fig:collisions:Trap_depth} shows the effect of $U_\mathrm{coll}$ on the atom loss at three different values of the magnetic field near observed loss features. We choose a hold time $t_{\rm coll}$ at each magnetic field such that there is only one inelastic collision on average; this is achieved by setting $t_{\rm coll}n_{2}k_{2}$ approximately equal to one.\footnote{One inelastic collision causes an increase in the kinetic energy of the atoms by at least an order of magnitude. The resulting decrease in the density makes a second inelastic collision highly unlikely for the chosen hold times.}
For low tweezer depths ($U_\textrm{coll} < \Delta E_1$) we observe a high value of $P_0$ for all three magnetic fields, indicating dominant $2 \to 0$ loss as expected. For increasing $U_{\rm coll}$, we see a decrease in $P_{0}$ and a commensurate increase in $P_{1}$ and $P_{2}$. The increase in $P_{1}$ precedes the increase in $P_{2}$. In Fig.~\ref{fig:collisions:Trap_depth}(d)-(f) we have rescaled the x-axis by $\Delta E_{1}$. This shows that, for magnetic fields of 101.5~G and 87.5~G, the centre of the rising edge of $P_{1}$ coincides with $\Delta E_{1}$ and the rising edge of $P_{2}$ coincides with $2\Delta E_{1}$. We note that the suppression of two-atom loss cannot be explained by the increase in collision energy $E_\mathrm{coll}$ with trap depth. For the largest tweezer depth, $U_\mathrm{coll}/h = 60$~MHz ($U_\mathrm{coll}/k_{\rm B} =2.9$~mK), we estimate the temperature of the atoms to be  $17(3)~\mu$K. At this increased collision energy, coupled-channel calculations indicate that $k_{2}$ is reduced by a factor of 15. The increased confinement leads to an increase in $n_{2}$ by a factor of around 30, so we expect an overall increase of inelastic collision rate with tweezer depth.

We compare our measurements with coupled-channel calculations that determine the dominant decay pathways at the magnetic fields investigated experimentally. 
The branching ratios of the dominant $\Delta M_F=+1$ and $+2$ decay pathways, including any resonant effects, are readily obtained from the off-diagonal elements of the S-matrices already calculated. 
We find that, for 87.5~G and 101.5~G, the loss is dominated by $\Delta M_F=+1$ by factors of 30 and 10, respectively, such that each atom receives energy $\Delta E_1$ from an inelastic collision. If the trap depth is between $\Delta E_1$ and $2\Delta E_1$, neither atom individually has enough kinetic energy to leave, but there is sufficient combined kinetic energy for \emph{one} atom to leave and the other to remain; this is observed as a sharp decrease in $P_0$ and corresponding increase in $P_1$. This suggests efficient kinetic energy transfer between the two atoms after the inelastic collision, which could occur through mechanisms such as elastic collisions in the collision tweezer or light-induced inelastic collisions in the imaging tweezers. As the trap depth increases, it is less likely for enough energy to be transferred to one atom for it to leave the trap, so $2\to 1$ loss is suppressed and $P_2$ increases. $P_1$ changes relatively little as $2\to 1$ loss from collisions is replaced by $2\to 1$ imaging loss. The likelihood of $2\to 1$ imaging loss increases relative to $2\to 0$ imaging loss as the kinetic energy of the atoms approaches the depth of the imaging tweezer \cite{Sompet2013}. 
For 32.8~G, the coupled-channel calculations indicate that the two possible decay pathways with $\Delta M_F=+1$ and $+2$ are approximately equally likely. This results in more energy released on average, shifting the transitions described above to larger trap depths. The existence of two competing pathways also broadens these features. 

\section{Conclusions}\label{sec:conclusion}

We have presented measurements of Feshbach resonances using pairs of Cs atoms prepared in a single optical tweezer. By performing inelastic loss spectroscopy of Cs pairs prepared in the $(f=3,m_f=-3)$ state, we have observed six loss features. The positions of the loss features are comparable to those observed in earlier experiments in bulk gases \cite{Leo2000, Chin2004}. However, by comparison with coupled-channel scattering calculation we have shown that the loss features are mostly centred on zeroes in the scattering length, rather than resonance centres. When the background scattering length is large, this effect can occur even at moderately low collision energies ($E_\textrm{coll}/k_\mathrm{B} \approx 2~\mu$K), which are typical for a tweezer-based experiment. By measuring the number of particles before and after the collision, we have also probed the rates of $2 \to 1$ and $2 \to 0$ atom loss as a function of tweezer depth. Comparison of these measurements with our calculations demonstrates that this is a viable technique for probing the energy released in an inelastic collision. Using radiative loss spectroscopy, we have also observed four Feshbach resonances for Cs pairs prepared in their lowest-energy state $(f=3,m_f=+3)$. Such resonances are normally detected via three-body loss in bulk gases. The centres and widths are in line with previously measured values \cite{Vuletic1999, Chin2004, Mark2018}. 

We have established the accuracy of Feshbach spectroscopy performed on atom pairs confined in optical tweezers by benchmarking our measurements against other measurements performed in bulk gases and our own coupled-channel calculations. Such tweezer-based measurements may be advantageous in systems where preparing high-density bulk gases is experimentally challenging. Collisions of laser-cooled CaF molecules have already been studied in optical tweezers \cite{Cheuk2020}, and the measurement techniques presented here may be extended to study Feshbach resonances between molecules \cite{Quemener2012} or between atoms and molecules \cite{Yang2019}. 

\section{Acknowledgements}
The authors thank A Alampounti for his assistance in developing software control of the AWG. This work was supported by U.K. Engineering and Physical Sciences Research Council (EPSRC) Grant EP/P01058X/1 and Durham University.

\section{Data Availability Statement}

The data that support the findings of this study are openly available at the following URL/DOI: \href{http://doi.org/10.15128/r2pr76f3410}{10.15128/r2pr76f3410}.

\vspace{20pt}

\bibliographystyle{iopart-num.bst}
\bibliography{references}

\providecommand{\newblock}{}
\begin{thebibliography}{10}
\expandafter\ifx\csname url\endcsname\relax
  \def\url#1{{\tt #1}}\fi
\expandafter\ifx\csname urlprefix\endcsname\relax\def\urlprefix{URL }\fi
\providecommand{\eprint}[2][]{\url{#2}}

\bibitem{Tuchendler2008}
Tuchendler C, Lance A~M, Browaeys A, Sortais Y~R~P and Grangier P 2008 {\em
  Phys. Rev. A\/} {\bf 78} 033425

\bibitem{Serwane2011}
Serwane F, Zürn G, Lompe T, Ottenstein T~B, Wenz A~N and Jochim S 2011 {\em
  Science\/} {\bf 332} 336--338

\bibitem{Xu2015}
Xu P, Yang J, Liu M, He X, Zeng Y, Wang K, Wang J, Papoular D~J, Shlyapnikov
  G~V and Zhan M 2015 {\em Nat. Commun.\/} {\bf 6} 7803

\bibitem{Liu2018}
Liu L~R, Hood J~D, Yu Y, Zhang J~T, Hutzler N~R, Rosenband T and Ni K~K 2018
  {\em Science\/} {\bf 360} 900--903

\bibitem{Sompet2019}
Sompet P, Szigeti S~S, Schwartz E, Bradley A~S and Andersen M~F 2019 {\em Nat.
  Commun.\/} {\bf 10} 1889

\bibitem{Cheuk2020}
Cheuk L~W, Anderegg L, Bao Y, Burchesky S, Yu S~S, Ketterle W, Ni K~K and Doyle
  J~M 2020 {\em Phys. Rev. Lett.\/} {\bf 125} 043401

\bibitem{Reynolds2020}
Reynolds L~A, Schwartz E, Ebling U, Weyland M, Brand J and Andersen M~F 2020
  {\em Phys. Rev. Lett.\/} {\bf 124} 073401

\bibitem{Brooks2021}
Brooks R~V, Spence S, Guttridge A, Alampounti A, Rakonjac A, McArd L~A, Hutson
  J~M and Cornish S~L 2021 {\em N. J. Phys.\/} {\bf 23} 065002

\bibitem{Chin2010}
Chin C, Grimm R, Julienne P and Tiesinga E 2010 {\em Rev. Mod. Phys.\/} {\bf
  82} 1225--1286

\bibitem{Leggett2001}
Leggett A~J 2001 {\em Rev. Mod. Phys.\/} {\bf 73} 307--356

\bibitem{Kohler2006}
K\"{o}hler T, G\'{o}ral K and Julienne P~S 2006 {\em Rev. Mod. Phys.\/} {\bf
  78} 1311--1361

\bibitem{Weber2003b}
Weber T, Herbig J, Mark M, N\"agerl H~C and Grimm R 2003 {\em Phys. Rev.
  Lett.\/} {\bf 91} 123201

\bibitem{Kraemer2006}
Kraemer T, Mark M, Waldburger P, Danzl J~G, Chin C, Engeser B, Lange A~D, Pilch
  K, Jaakkola A, N{\"a}gerl H~C and Grimm R 2006 {\em Nature\/} {\bf 440}
  315--318

\bibitem{Mark2018}
Mark M~J, Meinert F, Lauber K and Nägerl H~C 2018 {\em SciPost Phys.\/} {\bf
  5} 055

\bibitem{Jachymski2020}
Jachymski K 2020 {\em J. Phys. B\/} {\bf 53} 065302

\bibitem{Ketterle1996}
Ketterle W and Druten N~V 1996 {\em Adv. At. Mol. Opt. Phy.\/} {\bf 37}
  181--236 ISSN 1049-250X

\bibitem{Anderegg2019}
Anderegg L, Cheuk L~W, Bao Y, Burchesky S, Ketterle W, Ni K~K and Doyle J~M
  2019 {\em Science\/} {\bf 365} 1156--1158

\bibitem{Sala2013}
Sala S, Z\"urn G, Lompe T, Wenz A~N, Murmann S, Serwane F, Jochim S and Saenz A
  2013 {\em Phys. Rev. Lett.\/} {\bf 110} 203202

\bibitem{Hood2020}
Hood J~D, Yu Y, Lin Y~W, Zhang J~T, Wang K, Liu L~R, Gao B and Ni K~K 2020 {\em
  Phys. Rev. Res.\/} {\bf 2} 023108

\bibitem{Zhang2020}
Zhang J~T, Yu Y, Cairncross W~B, Wang K, Picard L~R~B, Hood J~D, Lin Y~W,
  Hutson J~M and Ni K~K 2020 {\em Phys. Rev. Lett.\/} {\bf 124}(25) 253401

\bibitem{Vuletic1999b}
Vuleti\ifmmode~\acute{c}\else \'{c}\fi{} V, Kerman A~J, Chin C and Chu S 1999
  {\em Phys. Rev. Lett.\/} {\bf 82} 1406--1409

\bibitem{Vuletic1999}
Vuleti\ifmmode~\acute{c}\else \'{c}\fi{} V, Chin C, Kerman A~J and Chu S 1999
  {\em Phys. Rev. Lett.\/} {\bf 83} 943--946

\bibitem{Chin2000}
Chin C, Vuleti\ifmmode~\acute{c}\else \'{c}\fi{} V, Kerman A~J and Chu S 2000
  {\em Phys. Rev. Lett.\/} {\bf 85} 2717--2720

\bibitem{Kerman2001}
Kerman A~J, Chin C, Vuleti\ifmmode~\acute{c}\else \'{c}\fi{} V, Chu S, Leo P~J,
  Williams C~J and Julienne P~S 2001 {\em C. R. Acad. Sci.\/} {\bf 2} 633--639

\bibitem{Chin2003}
Chin C, Kerman A~J, Vuleti\ifmmode~\acute{c}\else \'{c}\fi{} V and Chu S 2003
  {\em Phys. Rev. Lett.\/} {\bf 90} 033201

\bibitem{Herbig2003}
Herbig J, Kraemer T, Mark M, Weber T, Chin C, N{\"a}gerl H~C and Grimm R 2003
  {\em Science\/} {\bf 301} 1510--1513

\bibitem{Chin2004}
Chin C, Vuleti\ifmmode~\acute{c}\else \'{c}\fi{} V, Kerman A~J, Chu S, Tiesinga
  E, Leo P~J and Williams C~J 2004 {\em Phys. Rev. A\/} {\bf 70}(3) 032701

\bibitem{Koppinger2014b}
Köppinger M~P, Gregory P~D, Jenkin D~L, McCarron D~J, Marchant A~L and Cornish
  S~L 2014 {\em N. J. Phys.\/} ISSN 1367-2630

\bibitem{Spence2022}
Spence S, Brooks R~V, Ruttley D~K, Guttridge A and Cornish S~L 2022 {\em
  Unpublished\/}

\bibitem{Schlosser2002}
Schlosser N, Reymond G and Grangier P 2002 {\em Phys. Rev. Lett.\/} {\bf 89}
  023005

\bibitem{Endres2016}
Endres M, Bernien H, Keesling A, Levine H, Anschuetz E~R, Krajenbrink A, Senko
  C, Vuletic V, Greiner M and Lukin M~D 2016 {\em Science\/} {\bf 354}
  1024--1027

\bibitem{Liu2019}
Liu L~R, Hood J~D, Yu Y, Zhang J~T, Wang K, Lin Y~W, Rosenband T and Ni K~K
  2019 {\em Phys. Rev. X\/} {\bf 9}(2) 021039

\bibitem{Savard1997}
Savard T~A, O'Hara K~M and Thomas J~E 1997 {\em Phys. Rev. A\/} {\bf 56}(2)
  R1095--R1098

\bibitem{McGovern2011}
McGovern M, Hilliard A~J, Gr\"{u}nzweig T and Andersen M~F 2011 {\em Opt.
  Lett.\/} {\bf 36} 1041--1043

\bibitem{Jackson2020}
Jackson N~C, Hanley R~K, Hill M, Leroux F, Adams C~S and Jones M~P~A 2020 {\em
  SciPost Phys.\/} {\bf 8}(3) 38

\bibitem{Berninger2013}
Berninger M, Zenesini A, Huang B, Harm W, N\"agerl H~C, Ferlaino F, Grimm R,
  Julienne P~S and Hutson J~M 2013 {\em Phys. Rev. A\/} {\bf 87} 032517

\bibitem{Frye2019}
Frye M~D, Yang B~C and Hutson J~M 2019 {\em Phys. Rev. A\/} {\bf 100} 022702

\bibitem{Leo2000}
Leo P~J, Williams C~J and Julienne P~S 2000 {\em Phys. Rev. Lett.\/} {\bf
  85}(13) 2721--2724

\bibitem{Jones2006}
Jones K~M, Tiesinga E, Lett P~D and Julienne P~S 2006 {\em Rev. Mod. Phys.\/}
  {\bf 78}(2) 483--535

\bibitem{Fioretti1999}
{Fioretti, A}, {Comparat, D}, {Drag, C}, {Amiot, C}, {Dulieu, O},
  {Masnou-Seeuws, F} and {Pillet, P} 1999 {\em Eur. Phys. J. D\/} {\bf 5}
  389--403

\bibitem{Burnett1996}
Burnett K, Julienne P~S and Suominen K~A 1996 {\em Phys. Rev. Lett.\/} {\bf
  77}(8) 1416--1419

\bibitem{molscat:2019}
Hutson J~M and Le~Sueur C~R 2019 {\em Comp. Phys. Comm.\/} {\bf 241} 9--18

\bibitem{mbf-github:2020}
Hutson J~M and Le~Sueur C~R 2020 {\sc molscat}, {\sc bound} and {\sc field},
  version 2020.0 \url{https://github.com/molscat/molscat}

\bibitem{Manolopoulos:1986}
Manolopoulos D~E 1986 {\em J. Chem. Phys.\/} {\bf 85} 6425--6429

\bibitem{Alexander:1987}
Alexander M~H and Manolopoulos D~E 1987 {\em J. Chem. Phys.\/} {\bf 86}
  2044--2050

\bibitem{Hutson:res:2007}
Hutson J~M 2007 {\em New J. Phys.\/} {\bf 9} 152

\bibitem{Frye:resonance:2017}
Frye M~D and Hutson J~M 2017  {\bf 96} 042705

\bibitem{Sompet2013}
Sompet P, Carpentier A~V, Fung Y~H, McGovern M and Andersen M~F 2013 {\em Phys.
  Rev. A\/} {\bf 88}(5) 051401

\bibitem{Quemener2012}
Qu{\'e}m{\'e}ner G and Julienne P~S 2012 {\em Chem. Rev.\/} {\bf 112}
  4949--5011

\bibitem{Yang2019}
Yang H, Zhang D~C, Liu L, Liu Y~X, Nan J, Zhao B and Pan J~W 2019 {\em
  Science\/} {\bf 363} 261--264 ISSN 0036-8075

\end{thebibliography}

\end{document}